# Scalable and Versatile Linear Computation with Minimalistic Photonic Matrix Processor

*Zhaoang Deng[1,†], Zhenhua Li[1,†], Jie Liu[1,†,\*], Chuyao Bian[1], Jiaqing Li[1], Ranfeng Gan[1], Zihao Chen[1], Kaixuan Chen[2], Changjian Guo[2], Liu Liu[3,\*] and Siyuan Yu[1]*

[1]State Key Laboratory of Optoelectronic Materials and Technologies, School of Electronics and Information Technology, Sun Yat-Sen University, Guangzhou 510006, China.

[2]Guangdong Provincial Key Laboratory of Optical Information Materials and Technology, South China Academy of Advanced Optoelectronics, South China Normal University, Guangzhou 510006, China

[3] State Key Laboratory of Extreme Photonics and Instrumentation, College of Optical Science and Engineering, International Research Center for Advanced Photonics, Zhejiang University, Hangzhou 310058, China

[†] *These authors contributed equally.*

*\*Corresponding author: liujie47@mail.sysu.edu.cn; liuliuopt@zju.edu.cn*

## Abstract

The advancement of artificial intelligence demands flexible multimodal data processing with high throughput and energy efficiency. Photonic integrated circuits (PIC) has demonstrated promising potentials in terms of low latency and low power consumption per operation for linear operations such as matrix-vector multiplication. However, the existing schemes face challenges in their scalability due to the use of photonic circuits that expand with the scale of the operants, despite efforts of exploiting the multiple optical parameter dimensions such as time, wavelength and spatial parallelism. They also lacked flexibility and efficiency in switching between different types of operations or tasks and adapting to multimodal data. In this article, we introduce an optical matrix processor (MP) with a minimalistic recursive structure for both multiplications and accumulations. The MP consists of an eletro-optic ring-modulator implemented as a thin-film lithium niobate PIC that allows flexible configurability and time-division multiplexed scheduling. The MP supports not only versatile linear operations including vector/matrix-vector multiplication and single/multi-kernel convolution but also ultra-fast task switching and adaptability to data of different sizes, by simply adjusting the data baud rate relative to the ring delay without structural modifications. We demonstrate its capabilities in a optic-electronic convolutional neural network with a computing throughput up to 73.4 billion operations per second. The MP further supports high scalability through appropriate allocation of wavelength and space resources, extending computing parallelism to handle higher data volumes with higher energy efficiency. This novel scheme paves the way for a new class of photonic processors capable of managing escalating data workloads with unprecedented flexibility, efficiency and scalability.

## Introduction

Scaling up computational capacity of neuromorphic hardware using electronic integrated circuits faces practical challenges, including limitations in transistor counts and increased energy consumption for data movement[1,2]. In contrast, integrated photonics offer unique advantages, including low interconnect loss and vast bandwidth, as a promising platform for the development of optical computing accelerators[2-9].

Significant progress has been made in constructing integrated photonic neural networks, employing various on-chip schemes[10] to perform linear matrix operations in the photonics domain. These operations, including matrix multiplication and convolution, are essential computational steps in neural networks. However, existing schemes rely on the mapping of large-scale dense matrices on the spatial configurations of photonic integrated circuits (PIC), an approach that faces substantial limitations due to the large sizes of photonic compomnents hence limited scalability of PICs. Mapping a matrix of size $N \times N$ solely based on space division multiplexing (SDM) schemes typically require $O(N^2)$ scaling of photonic components, leading to very large footprint and increased complexity in fabrication and manipulation[11].

Much effort has been invested in reducing the count of photonic components in order to improve the computational scalability of photonic neural networks, mostly by multiplexing schemes exploiting the multiple parameter dimensions of the light wave. A linearly scaling compact integrated diffractive optical network (IDNN) that required $2N$ units to accommodate an input matrix of dimension $N$ was implemented by allowing 2D diffraction that exploits on-chip spatial cross-propagation of light[12]. In a subsequent publication[13], a wavelength division multiplexing (WDM) scheme were used to reduce the component count to $N$ units for an $N$-dimensional input matrix.

Nevertheless, the scalability of on-chip optical matrix operations essentially remains contingent on the scalability of photonic integration, calling for more efficient strategies to improve the scaling potential. In addition, existing matrix mapping processes often lacked flexibility, often requiring the reconfiguration of photonic architectures to implement different types of computations, such as unitary/nonunitary matrix multiplications[5,14] and/or convolutions[12,13,15,16]. The speed of weight updates is often not sufficiently rapid, rendering the system less effective for neural networks that demand online training or amalgamation of training and inference processes[17]. Innovative schemes are therefore desirable to overcome these limitations and fully harness the potentials of photonic neural networks.

In this paper, we propose a scalable and flexible on-chip optical matrix processing unit (MP) based on a recursive time-domain scheme implemented as a single thin-film lithium niobate (TFLN) ring-modulator. Leveraging the high modulation speed and the fast reconfigurable feedback loop of the TFLN ring-modulator, a single device can facilitate linear multiply-and-accumulate (MAC) operations necessary for the vector multiplications (i.e., dot-products). Comparing to the cascading MZM modulation schemes[18-23] which need additional optical or electrical conbiner/integrator to accumulate the element-wise products, the proposed minimalistic MP executes both multiplications and accumulations in optical domain by the recursive strategy, without the need for additional electronic or photonic components. By tuning the ratio between the modulation symbol period and the feedback loop's time delay, our MP design allows for the versatile expansion to matrix-vector multiplications. Furthermore, by strategically arranging data symbols in the time domain, we can further convert vector/matrix-vector multiplications into convolution operations. These conversions allows both vector/matrix-vector multiplication and single/multi-kernel convolutions completed in each computation cycle, and without requiring any changes to the existing on-chip architecture. Owing to the high-speed electro-optic interfaces of the MP, the time-division multiplexing (TDM)-based signal scheduling supports rapid weight updates and flexible matrix mapping, potentially adapting to multimodal data and computation tasks. To the best of our knowledge, this marks the first instance of achieving both matrix-vector multiplication and convolution with a single photonic device.

An optical convolution layer based on the MP in conjunction with an electrical fully connected layer forms a convolutional neural network (CNN). This is utilized to perform a ten-class classification task on the Modified National Institute of Standards and Technology (MNIST)[24] dataset of handwritten digits, achieving an accuracy of 88.3%. Additionally, the same MP demonstrates feature extractions that range from single-kernel to multi-kernel convolution within a single computational cycle, maintaining a high accuracy rate of 88.7%. These results highlight the MP's scalability and versatility, without modifications on the minimalistic photonic architecture.

The propose MP provides an attractive strategy for computational frameworks that require rapid parameter updates or the ability to adaptively switch among various operations, including the Ising machines[25,26], distributed AI systems[20,27] and adaptive compensation of communication channel impairments[28,29]. Furthermore, the MP also maintain the potential to extent computing scalability and throughput by integrating with SDM and WDM strategies. Such architectures integrating time, wavelength, and spatial domain resources has the potential to underscore a new paradigm for future optical computing frameworks.

In the subsequent parts of the article, the Section **Operating Principle** presents the principle for basic vector multiplication, followed by versatile extensions on matrix-vector multiplication and single/multi-kernel convolution. Next, the Section **Experimental Demonstrations** introduces the ring-modulator and the detail demonstration on feature extractions and CNNs based on the MP's capability for single/multi-kernel convolution. Then, the Section **Discussions** explores the application scenarios, performance metrics, and scaling strategies of the MP. Finally, the Section **Conclusions** summarizes the entire article.

# Results

## Operating Principle

Fig.1 illustrates the structural diagram of the proposed MP. The primary component of the MP, a high-speed electro-optic ring-modulator, comprises a standard 2 × 2 Mach–Zehnder interferometer (MZI) and a loop waveguide linking the MZI's two ports[30]. The input data is encoded onto optical light via an external intensity electro-optic modulator and then fed to the MP for executing MAC operations necessary for vector multiplication, matrix-vector multiplication and convolution. This section will detail the operating principles of vector multiplication using a ring-modulator, as well as the versatile extension strategies for vector-matrix multiplication, single/multi-kernel convolution, preserving the existing on-chip photonic architecture.

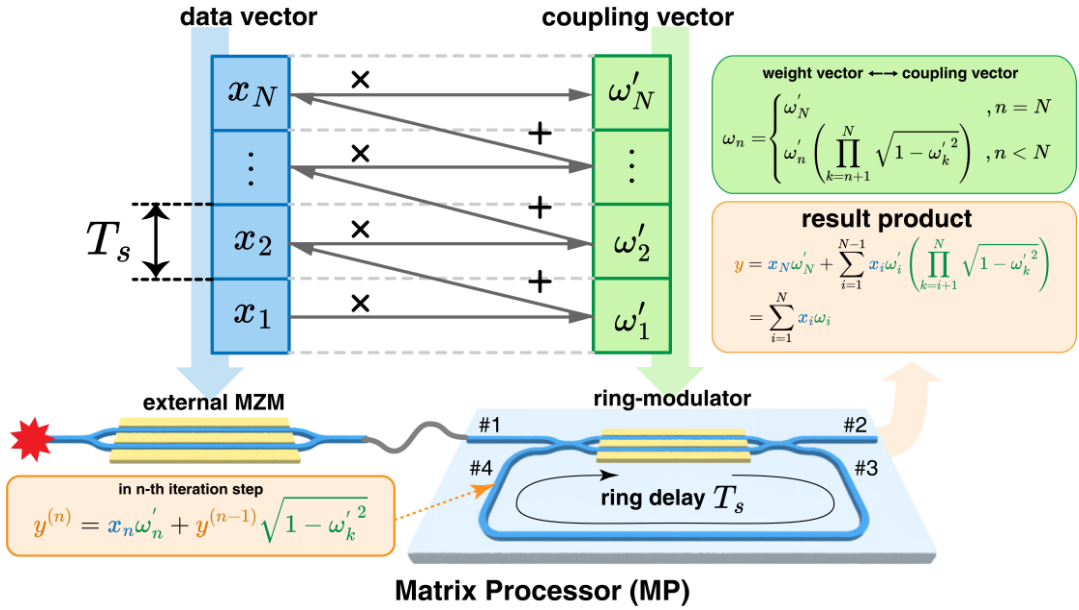

Fig.1. Vector multiplication based on the proposed MP.

**Vector Multiplication Based on the MP**

As illustrated in Fig.1, an optical carrier is serially modulated with elements of the data vector $X = [x_1, x_2, \cdots, x_N]^T$ through an external Mach–Zehnder modulator (MZM), each element occupying a symbol period of $T_s$ that is the recursive delay of the ring. Subsequently, the generated data signal passes through the ring-modulator for MAC operations. Simultaneously, the weight vector $W = [\omega_1, \omega_2, \cdots, \omega_N]^T$ is initially transformed into the coupling vector $W' = [\omega'_1, \omega'_2, \cdots, \omega'_N]^T$ to drive the integrated MZI of the ring-modulator at a symbol period of $T_s$ as well. It is important to note that the relationships between vectors $W$ and $W'$ can be found in Supplementary Note 1.

The multiplication operation for data and weight symbols can be achieved by precisely aligning $X$ with the corresponding element of $W'$ in the time domain. Subsequently, the newly generated symbol after the multiplication operation at port #3 of the MZI (as shown in Fig.1) follows the loop waveguide of the ring-modulator back to port #4 with a delay time $T_s$ equal to one symbol period, and then undergoes accumulation by coherent optical superposition with the next data symbol injected at port #1. This process is defined as one iteration step of the MAC operations for vector multiplication in the proposed MP. Therefore, after the $N^{th}$ iteration step, the generated signal from port #2 can be expressed as (the detail can be found in Supplementary Note 1):

$$y = x_N \omega'_N + \sum_{i=1}^{N-1} x_i \omega'_i \left( \prod_{k=i+1}^{N} \sqrt{1 - {\omega'_k}^2} \right). \tag{1}$$

Following the the relationships between vectors $W$ and $W'$, Eq.1 can be rewritten as the vector multiplication of data vector $X$ and weight vector $W$:

$$y = \sum_{i=1}^{N} x_i \omega_i, \tag{2}$$

whose intensity is then detected by the photodetector (PD) to generate the photocurrent:

$$I = \mu |y|^2, \tag{3}$$

where $\mu$ is the PD responsivity. Thus, directly applying square root and normalization on $I$, we can get the vector multilication result $y = |y| = \sqrt{I/\mu}$ when $y \geq 0$. To avoid negative $y$, the arbitrary real weight matrix could be splitted into two positive part to garuantee positive results, which is discussed in Supplementary Note 2.

**Matrix-Vector Multiplication Based on the MP**

As shown in Fig.2, the M × N data matrix is firstly vectorized into a MN-dimension data vector $X$, connecting each end of column to the start of next, as follows:

$$\begin{bmatrix} x_{11} & \cdots & x_{1N} \\ \vdots & \ddots & \vdots \\ x_{M1} & \cdots & x_{MN} \end{bmatrix} \Rightarrow X = [x_{11}, x_{21}, \cdots, x_{M1}, \cdots, x_{1N}, \cdots, x_{MN}]^T, \tag{4}$$

and then encoded as optical signal by the external MZM. Simultaneously, the weight vector $W$ is transformed to be coupling vector $W'$, then loaded on the ring-modulator for MAC operations of the matrix-vector multiplication.

To achieve a matrix-vector multiplication, the key distinction from the vector multiplication is that the symbol period of $X$ is set to be $T_s/M$, but that of $W'$ maintains $T_s$, as illustrated in Fig.2. Therefore, during one iteration step, each element of $W'$ is possible to interact with M symbols of $X$ in the ring. Consequently, with a computation cycle with N iteration steps, the MP facilitates the N-dimension vector multiplications for M times—equivalently, a matrix-vector multiplication with an M × N data matrix and an N-dimension weight vector. The outcomes of this matrix-vector multiplication are presented as an M-dimension vector $Y = [y_1, y_2, \ldots, y_M]^T$

with entries $y_i$ expressed as:

$$y_i = \sum_{j=1}^{N} x_{ij}\omega_j. \tag{5}$$

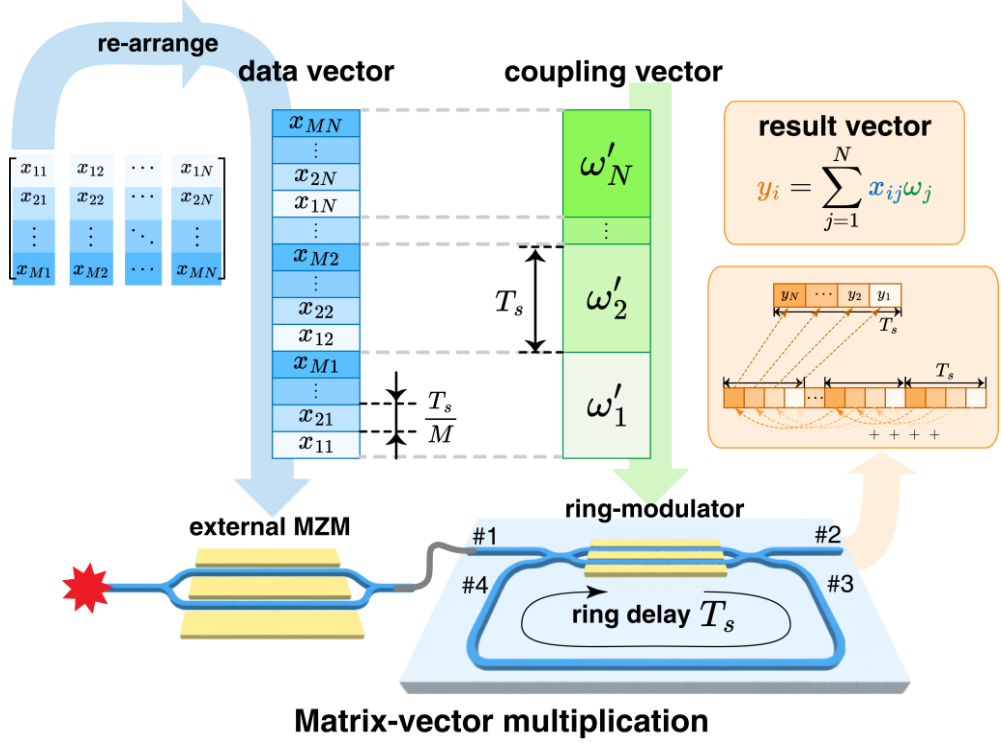

Fig.2 The process of M×N matrix multiplication in MP.

**Convolution Operations with Single Kernel**

The strategic re-arrangement of matrix elements enables the operation swithing from the matrix-vector multiplication to convolution operations using the proposed MP. Here we consider the example of convolving a 4×4 data matrix with a 2×2 convolution kernel as followed:

$$\begin{bmatrix} x_{11} & x_{12} & x_{13} & x_{14} \\ x_{21} & x_{22} & x_{23} & x_{24} \\ x_{31} & x_{32} & x_{33} & x_{34} \\ x_{41} & x_{42} & x_{43} & x_{44} \end{bmatrix} * \begin{bmatrix} \omega_{11} & \omega_{12} \\ \omega_{21} & \omega_{22} \end{bmatrix} \Leftrightarrow \begin{bmatrix} x_{11} & x_{21} & x_{12} & x_{22} \\ x_{12} & x_{22} & x_{13} & x_{23} \\ \vdots & \vdots & \vdots & \vdots \\ x_{33} & x_{43} & x_{34} & x_{44} \end{bmatrix} \begin{bmatrix} \omega_{11} \\ \omega_{21} \\ \omega_{12} \\ \omega_{22} \end{bmatrix}, \tag{6}$$

where $*$ is denoted as convolution.

According to Eq.6, the 4×4 data matrix is reorganized according to convolution rules into a 9×4 matrix. Simultaneously, the 2×2 convolution kernel is reshaped into a 4×1 vector, facilitating the necessary MAC operations for the convolution process. Consequently, this convolution equates to multiplication with a 9×4 matrix and a 4×1 vector. Following calculations adhere to the procedures outlined in the preceding section.

**Convolution Operations with Multiple Kernels**

The proposed MP can further perform multi-kernel convolution operations. Similar with the conversion outlined in Eq.6, the convolueions involving a 4×4 data matrix and two 2×2 convolution kernels can be converted as follows:

$$\begin{bmatrix} x_{11} & x_{12} & x_{13} & x_{14} \\ x_{21} & x_{22} & x_{23} & x_{24} \\ x_{31} & x_{32} & x_{33} & x_{34} \\ x_{41} & x_{42} & x_{43} & x_{44} \end{bmatrix} * \begin{matrix} \begin{bmatrix} \omega_{1,11} & \omega_{1,12} \\ \omega_{1,21} & \omega_{1,22} \end{bmatrix} \\ \begin{bmatrix} \omega_{2,11} & \omega_{2,12} \\ \omega_{2,21} & \omega_{2,22} \end{bmatrix} \end{matrix} \Leftrightarrow \begin{bmatrix} x_{11} & x_{21} & x_{12} & x_{22} \\ x_{12} & x_{22} & x_{13} & x_{23} \\ \vdots & \vdots & \vdots & \vdots \\ x_{33} & x_{43} & x_{34} & x_{44} \end{bmatrix} \begin{bmatrix} \omega_{1,11} & \omega_{2,11} \\ \omega_{1,21} & \omega_{2,21} \\ \omega_{1,12} & \omega_{2,12} \\ \omega_{1,22} & \omega_{2,22} \end{bmatrix}.$$

(7)

Based on Eq.7, we can optionally complete the convolution for the first kernel following the the procedures outlined in the preceding section, before progressing to the next one. However, this sequential approach necessitates the removal of redundant optical energy in the ring after the first kernel computation (refer to Supplementary Note 3), leading to additional computational delay. To address this, an alternative strategy is proposed.

As depicted in Fig.3, the multi-kernel convolutions of Eq.7 are possibly achieved within a single computational cycle, by carefully managing the relationship between the symbol rates of signals generated by the external MZM and the ring-modulator. Based on the single-kernel convolution (or equivalently the matrix-vector multiplication shown in Fig.2), the kernel entries can be devided as M halves and interleaved as a single coupling vector, with symbol period of $T_s/MK$, or 1/K of that of the data vector, so that each data symbol corresponds to K kernel symbols from different K kernels. Accordingly, a single data symbol corresponds to K kernel symbols from K different kernels, and MK MAC results are generated in each of N iteration steps. After a computation cycle with N iteration steps, K resulted feature maps, each of which contains M pixels, are simultaneously ouputed. Compared to the successive single-kernel convolutions for each kernels, this kernel-interleaving approach maximizes the benefits of the high modulation speed of the THLN devices, and eliminates the need for additional iteration steps to reset the MP (clear energy out of the ring) between the successive computational cycles.

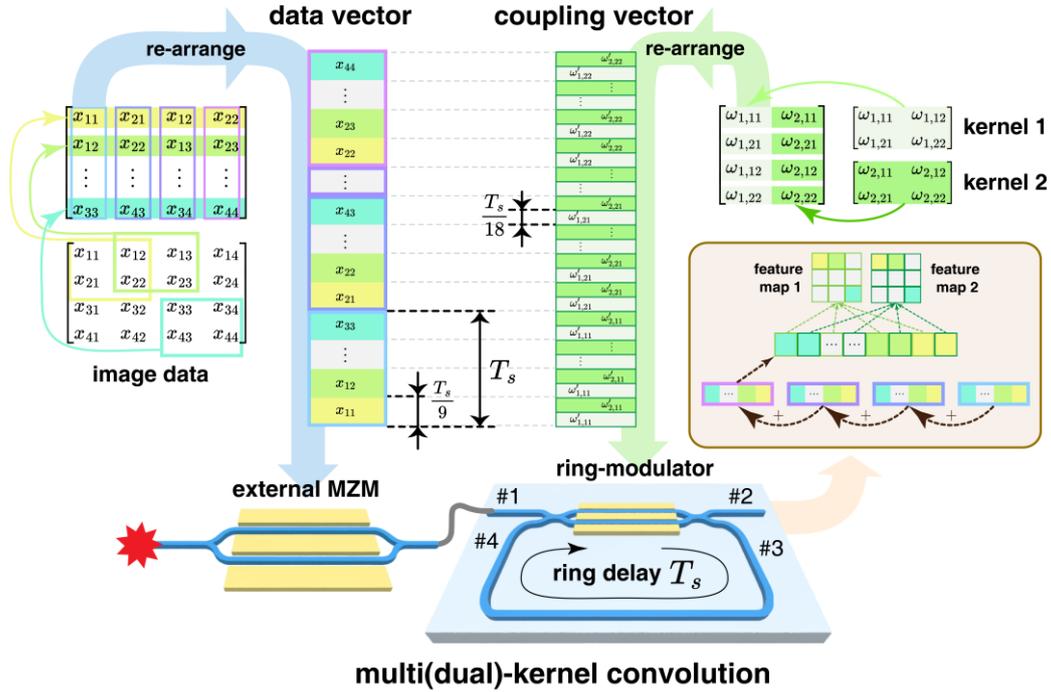

Fig.3 The process of multi(dual)-kernel convolution in MP.

Specifically, for a dual-kernel convolution (K = 2, M = 9, N = 4) as shown in Fig.3, the symbol rate of the kernels is double that of the data, and two kernels can interacting with distinct segments of the data symbol (e.g., the first and second halves) in each iteration steps, resulting in two 9-pixel feature maps within 4 iteration steps. In practice, under the constraint of modulation speed of the modulator, M could be smaller than the pixel number of feature map, thus several computation cycles should be executed to complete the entire image convolution.

## Experimental Demonstrations

To verify the feasibility of the proposed MP, a TFLN-based ring-modulator is fabricated to perform convolution operations. Here noted that the convolution operation is selected as a representative demonstration because, according to the theoretical framework, successfully completing convolutions indicates that the MP possesses the capability for vector/matrix-vector multiplication. By integrating the MP with an electrically fully connected layer, a synergistic optic-electronic CNN can be established. This CNN is employed to execute a MNIST handwritten digit classification task[24]. To demonstrate the scalability and versatility of the MP, scenarios involving both single and dual-kernel convolutions are presented.

### The TFLN-Based ring-modulator

The TFLN-based ring-modulator depicted in Fig.4(a) and (b), a key component of the MP, employing periodic capacitively loaded traveling-wave (CLTW) electrodes[30-33] to enhance the modulation bandwidth much higher than 67 GHz[30]. The device features a standard 2×2 MZI cooperating with a loop waveguide connecting two MZI ports. The half-wave voltage ($V_\pi$) of the MZI of about 7.5 V is obtained by conducting a sweep of DC bias and then reading the half of the range between two zero-coupling points, as shown in Fig.4(c). From the transmission spectrum shown in Fig.4(d), the free spectral range (FSR) of the ring is 18.35GHz near the wavelength of 1554 nm, i.e., time delay of the ring waveguide is 54.5 ps. Further details on the device are available in Section Materials and Methods and Supplementary Note 4.

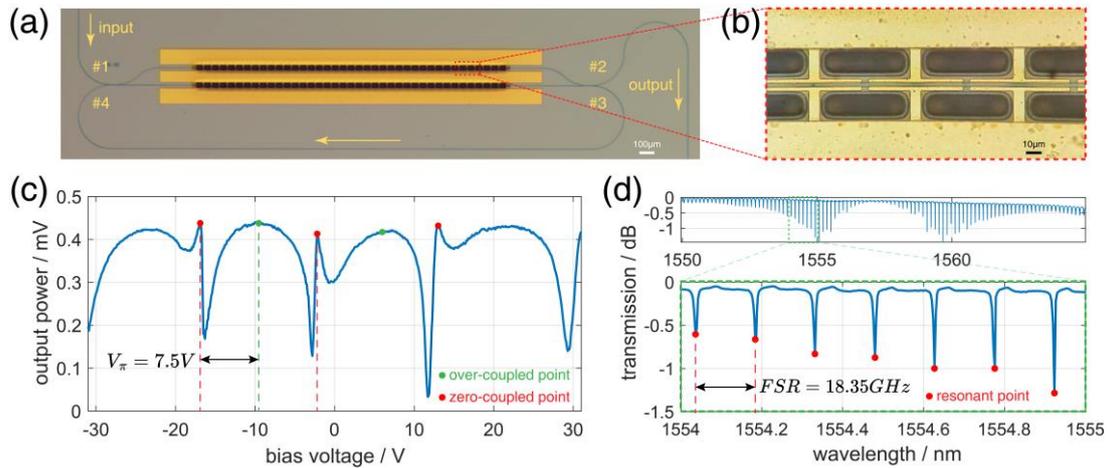

Fig.4 (a) The micrograph of the ring-modulator and (b) the partial zoom-in micrograph of the CLTW electrode. (c) The measured electro-optic transferring curve and (d) the transmission spectrum of the ring-modulator.

**Feature Extraction Based on Convolution Operations of the MP**

Fig.5 illustrates the experimental setup of the synergistic optic-electronic CNN, which consists two parts: (i) the optical convolution layer, to extract the feature maps of handwritten digits, and (ii) the electrical fully connected layer, to process the extracted feature maps to produce the predicted results. Details of the setup are introduced in Section Metarials and Methods.

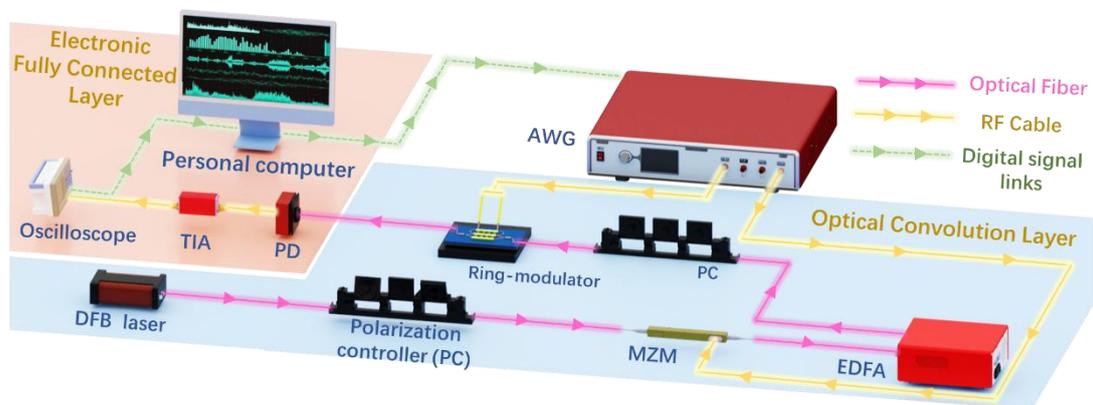

Fig.5 The Experiment Setup.

Before implementing the CNN, we firtly verified the convolution performance of the optical layer based on the proposed MP. As a representative example, the feature map extraction of a 28×28-pixel handwritten digit "5" (Fig.6(a)) was conducted, employing a selected convolution kernel $\begin{bmatrix} 0 & 1 \\ 0 & -1 \end{bmatrix}$. As shown in Fig.6, the process begins by rearranging the 28×28 image pixels into a 729 × 4 matrix, where 729 (27×27) denotes the count of convolution operations (as well as the number of pixels of a feature map) performed on the image using a 2×2 kernel (refer to Eq.6). This data matrix is then vectorized to serially load on the carrier by the external MZM at an 18.35 Gbaud/s data rate, as the waveform shown in Fig.6(b) and (c). This rate is specifically selected to synchronize with the time delay of the ring in the ring-modulator.

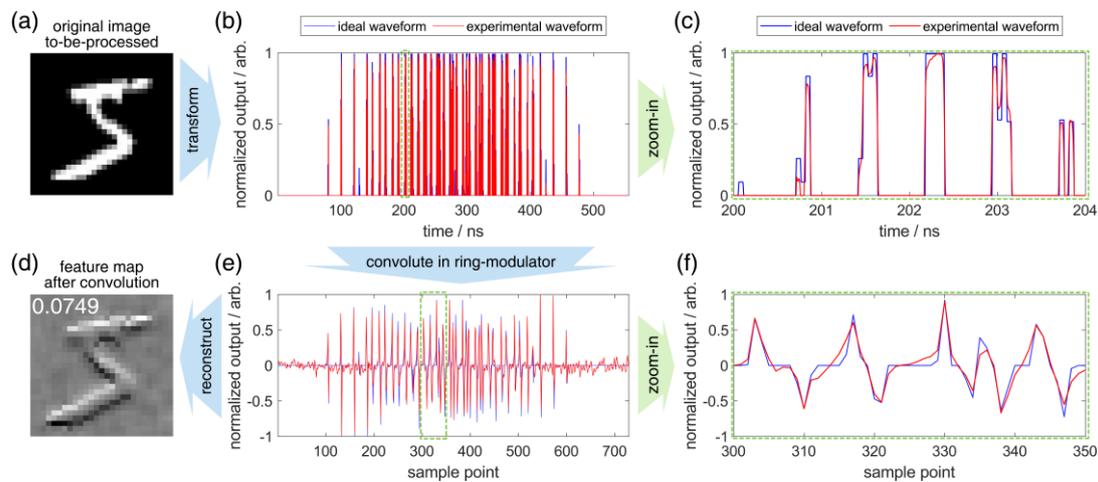

Fig.6 Feature map extraction. (a) The original image. (b) computer-generated ideal waveform (blue line) and the experimentally modulated waveform (red line) of the transformed pixels sequence. (c) The zoom-in waveform of the green dashed region of (b). (d) The resulted feature map after MP convolution. The RMSE is marked at the top-left. (e) The ideal waveform (blue line) and the experimental MP processed waveform (red line) of the convolution results. (f) The zoom-in waveform of the green dashed region of (e).

Subsequently, the optical image data signal is coupled into the ring-modulator. Here, it interacts with another electrical signal that encodes the convolution kernel information. This electrical signal carries a reshaped 4×1 vector from the kernel also at a symbol rate of 18.35 Gbaud/s. According to the convolution procedures detailed in the Section Convolution Operations with Single Kernel, each step of the convolution process involves multiplying a single row from the 729 × 4 data matrix with a 4×1 kernel vector. This procedure is repeated for all 729 rows to complete the required convolution. It is important to note that the bandwidth constraints of RF devices, such as, the external MZM, the RF amplifier, and PD, permit only a single optical symbol in the ring's loop waveguide per iteration step (i.e., $M = 1$) in the experiments. Consequently, despite the convolution could be converted to a matrix-vector multiplication, a series of vector multiplications (refer to Fig.1) with several computation cycles, rather than a single matrix-vector multiplication (refer to Fig.2) within one cycle, are performed to complete the required operations in the following experiments.

Fig.6(e), (f) and (d) displays the experimental waveform and reconstructed feature map obtained from the MP's convolution, despite the presence of distortion. The distortion has been partly mitigated using the methods detailed in Supplementary Note 5 and 6, leading to a root mean square error (RMSE, its calculation can be found in detail in Supplementary Note 7) of 0.0749 relative to the digital convolution results.

To demonstrate the general performance of the proposed MP, five MNIST images are randomly selected and convolved with 4 independent kernels to generate totally 20 feature maps, as shown in Fig.7, yielding an average RMSE of 0.0699 relative to the digital convolution groud-truth (listed in Table.1) and featuring 4.84 effective bit of precision (calculation details can be found in Supplementary Note 7). In addition, the MP a significantly higher speed than the digital method under the Intel Core i9-10980 CPU, as outlined in Table.1 (with detailed evaluation in Supplementary Note 8). The futher improvement of computational speed can be explored in Section Discussion.

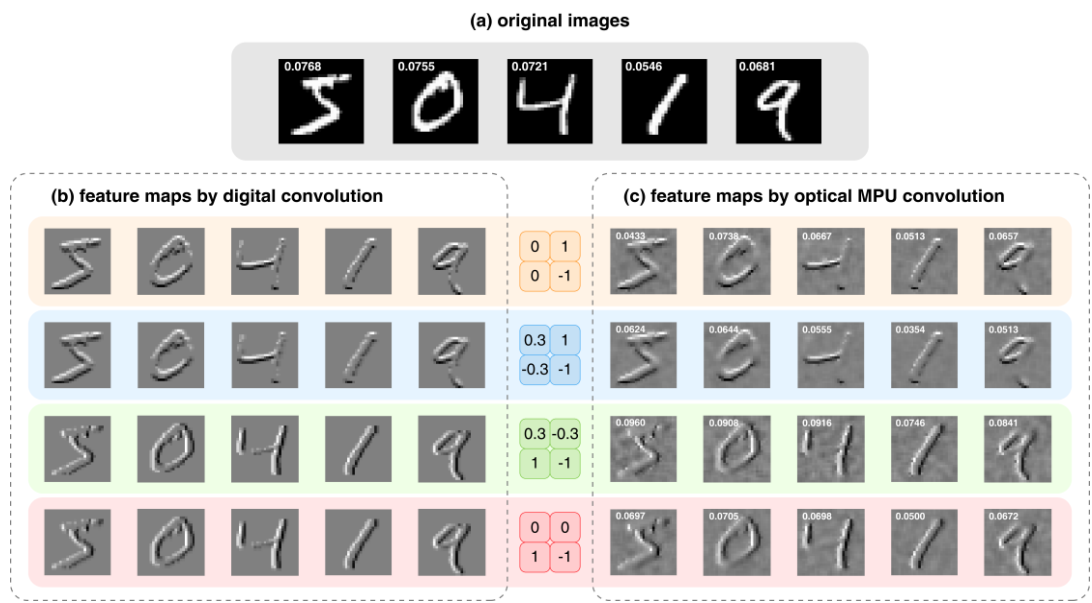

Fig.7 (a) Original images of "5", "0", "4", "1", and "9", with average RMSE regarding to each image marked at the top-left. Feature maps resulted by (b) digital convolution and (c) optical MP convolution, with RMSE marked at the top-left.

**Table.1 Performance comparison between the digital computer and MP on single-kernel convolution**

| Criteria | RMSE | Image processing speed (million Image/s) |
|---|---|---|
| Digital Computer (Intel Core i9-10980 CPU) | / | **0.14** |
| MP | **0.0699**<br>(in average across 20 feature maps) | **2.52** |

## CNN Classification Based on MP Extracted Feature Maps

To intuitively assess the effectiveness of feature extraction, the MP-based optical layer is combined with an electronic fully connected layer, which employs a linear rectification function (ReLU) for nonlinear activation[34] to construct a CNN for classifying handwritten digits, as illustrated in Fig.8(a). Within the MP-based optical convolution layer, two kernels independently generate two feature maps. It is important to note that two separate single-kernel convolution cycles are utilized for the two distinct feature maps, respectively. Following activation with ReLU, the two feature maps are merged into a feature vector, which is then input into the fully connected layer for classification. The electronic fully connected layer trained through offline stochastic gradient descent algorithms with adam optimizer to reduce cross-entropy loss[24]. The output vector indicates the predicted digit category, identified by the position of the maximum value within the vector.

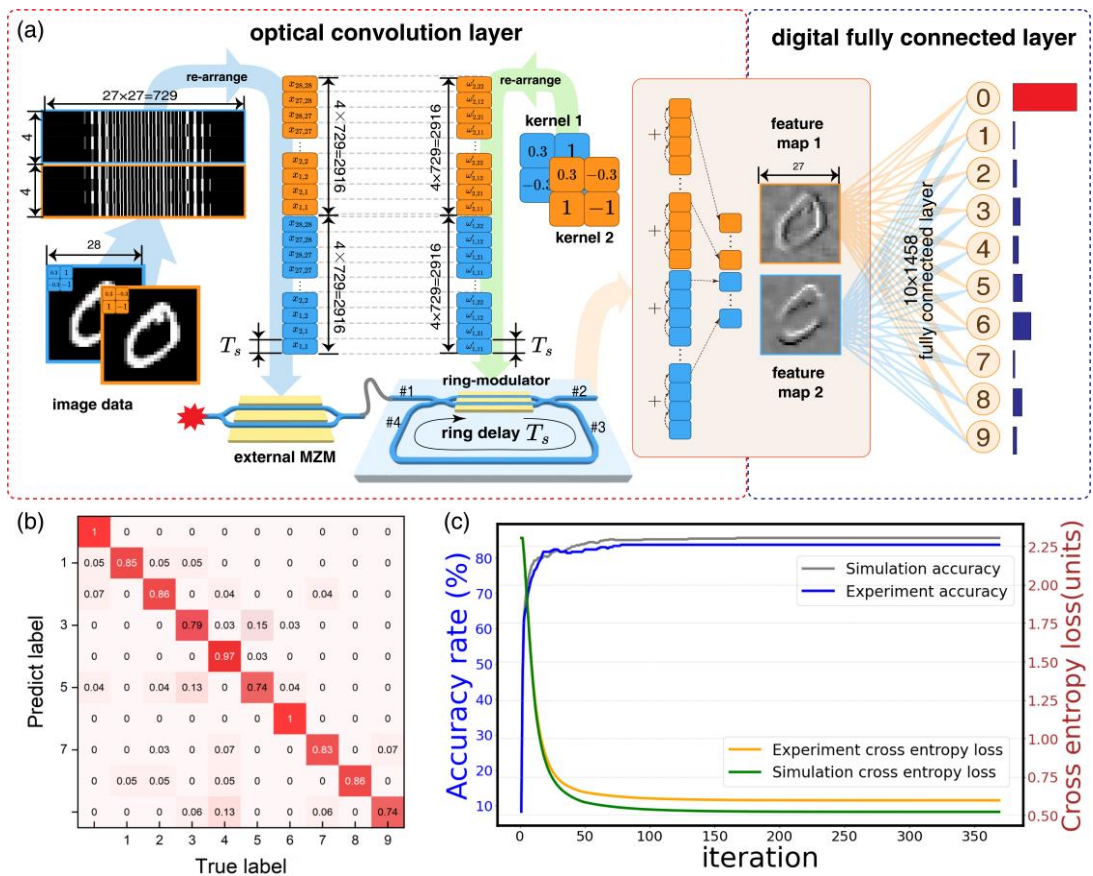

Fig.8 Classification result of MNIST handwritten digit image assisted by single-kernel convolution on MP. (a) The procudure of the demonstrated CNN, which contains an optical convolution layer for single-kernel convolution, and an electrically fully connected layer. (b) The confusion matrix of the experimental result by the CNN. (c) The comparison of experiment and simulation results during training.

An extensive evaluation on 2,000 images for classification is conducted, with 1,500 images used for training and 500 for testing. As shown in Fig.8(c), after 100 trainning iterations, the classification accuracy reaches a plateau, showing a consistent 3% discrepancy between the outcomes based on MP convolutions and digital convolutions (88.3% vs 91.3%).

**Scaling Up to Multi-Kernel Feature Extraction in Single Computation Cycle**

In order to underscore the computing scalability and versatility of the minimalistic MP, the opto-electronic CNN with an optical layer executing multi(dual)-kernel convolution on MP have demonstrared, as the schematic diagram shown in Fig.9(a). Without additional architectural modifications, two feature maps from two kernels can be extracted within a single computational cycle by simply setting the appropriate symbol period.

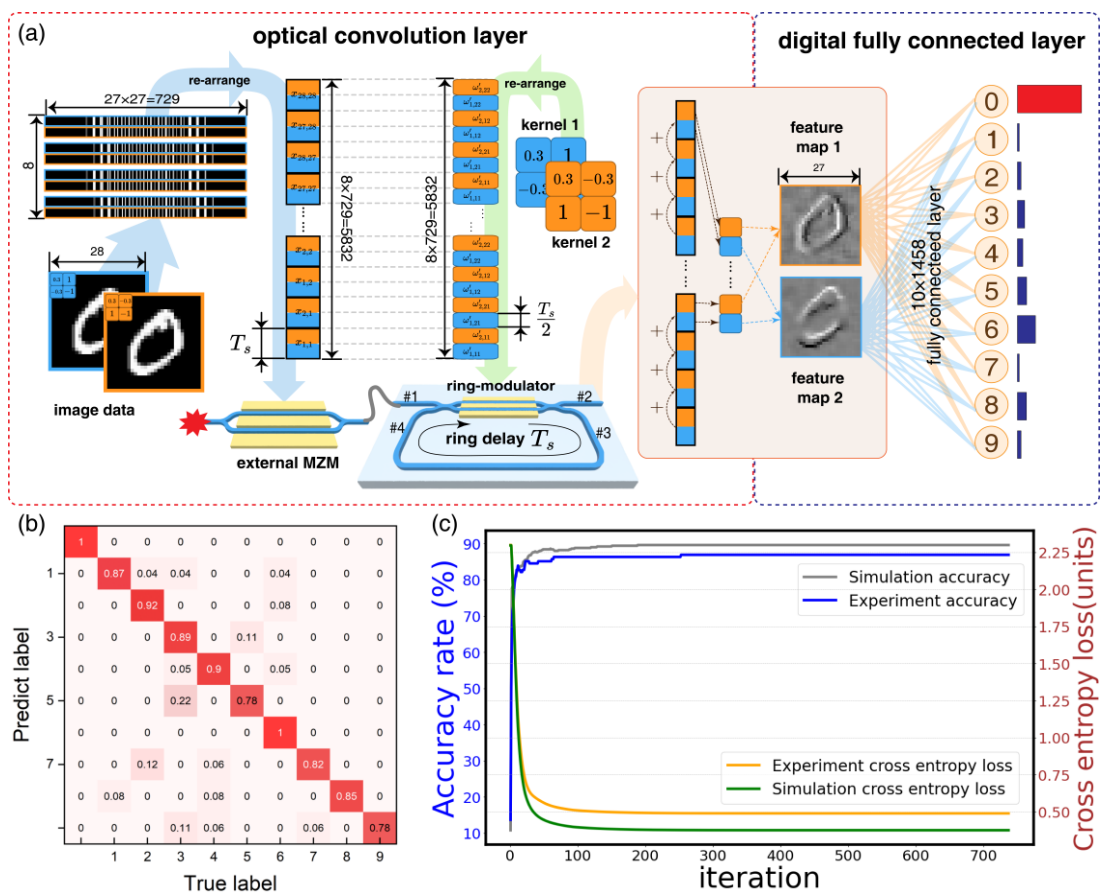

Fig.9 Classification result of MNIST handwritten digit image assisted by dual-kernel convolution on MP. (a) The procedure of the demonstrated CNN, which contains an optical convolution layer for dual-kernel convolution, and an electrically fully connected layer. (b) The confusion matrix of the experimental result by the CNN. (c) The comparison of experiment and simulation results during training.

As the schematic diagram shown in Fig.9(a), in the optical convolution layer, the optical image data signal from the external modulator is set to 18.35 Gbaud/s, equal to the time delay of the ring of the ring-modulator. Concurrently, the kernel signal which carries two convolution kernel vectors in a serially interleaved fashion, is fixed at 36.7 Gbaud/s, twice of that for the image data signal. Consequently, each data symbol processed by the ring-modulator underwent two distinct weighted MAC operations in its first and second halves in each iteration steps. Upon completing all iteration steps, this approach facilitated the extraction of features based on both kernels, producing a feature vector interleaved by two feature maps, as illustrated in Fig.9(a) (for the specific calculation process, refer to Section Convolution Operations with Multiple Kernels).

Randomly choosing 5 images, their resulted 10 feature maps show an average RMSE of 0.1494 relative to the digital groud-truth (bit precision of 3.74 bit equivalently), as listed in Table.2. This performance is degraded compared to that observed in the single-kernel feature extraction scenario (refer to Table 1). This degradation is primarily due to the bandwidth limitations of RF devices, such as RF probes, PDs, and RF amplifiers, which distort output symbols from the MP at higher modulation speeds. However, the higher modulation speed brings an higher image processing speed, as outlined in Table.2.

Table.2 Performance comparison between the digital computer and MP on dual-kernel convolution

| Criteria | RMSE | Image processing speed (million Image/s) |
| --- | --- | --- |
| Digital Computer (Intel Core i9-10980 CPU) | / | **0.14** |
| MP | **0.1494** (in average across 10 feature maps) | **5.04** |

Continuing to feed the feature vector into the electronic fully connected layer, the final prediction results are generated with classification accuracies of 88.7%, despite the relatively higher RMSE. This accuracy of 88.7% is only a 3.6% disparity compared to the digital CNN and is comparable to that of the single-kernel convolution, highlighting the computing scalability and versatility of the minimalist MP.

# Discussions

## Optimal Application Scenario for the MP

The proposed MP exhibits remarkably scalability and flexibility, facilitating seamless transitions between vector multiplication, matrix-vector multiplication, and convolution operations, all without the need for hardware modifications. This adaptability extends to the computational parameters within each operation type, offering versatility in handling tasks such as matrix-vector multiplication across diverse matrix forms, including unitary and non-unitary matrices. It also accommodates the use of single or multiple convolution kernels within a single convolution step. Moreover, by harnessing the high modulation speed and swift re-configurability of the electro-optic modulator, the proposed MP offers clear benefits for computational frameworks that necessitate rapid parameter updates. This level of versatility and fast re-configurability significantly enhances the MP's utility for applications requiring dynamic adaptability to varying solution requirements, such as the Ising machines[25,26], distributed AI systems[20,27] and communication channel equalization[28,29].

## Performance Comparisons

Table 3 presents a performance comparison of the proposed MP with typical integrated photonic matrix computation hardware. The MP is distinguished by its remarkable scalability and flexibility, which significantly increase its versatility. This enables seamless transitions between vector and matrix-vector multiplications, as well as between single and multi-kernel convolution operations, without necessitating modifications to the existing photonic layout on the chip. As listed in Table 3, our MP is the first demonstration with a single on-chip device to achieve both vector/matrix-vector multiplication and convolution. The advantages of the MP primarily stem from the high modulation speed and rapid reconfigurability of the TFLN ring-modulator within a time-division multiplexing (TDM) framework. However, these benefits may come at the cost of increased energy consumption and latency.

In the MP, each MAC operation requires electro-optic modulation, and operations are structured using a TDM scheme. This arrangement leads to frequent opto-electronic conversions and memory access, potentially increasing the energy demands of individual computations[35]. However, the MP's adaptable computational architecture mitigates this by allowing for the reuse of electro-optic conversion parameters. For instance, in convolutions derived from matrix-vector multiplications, the same convolution kernel can be employed across various data sets. Similarly, in operations involving multiple kernels, the kernels can process the same data set multiple times. This efficiency significantly lowers the frequency of memory access required for each operation, thus conserving energy. Consequently, the MP maintains an energy consumption of several pJ per operation, as detailed in Table 3. Furthermore, by strategically integrating wavelength-division and space-division multiplexing techniques, the computational scale of the MP can be further expanded (illustrated in Fig.10). This approach ensures high throughput and flexibility of MP while further reducing the electro-optic conversions required for each MAC operation, thus optimizing overall performance of power efficiency and versatility, as summarized in the last row of Table 3. More details about the metrics in Table 3 are provided in Supplementary Note 8.

Table 3 Comparative analysis of state-of-the-art integrated photonic computing hardware

| Type | Number of programmable elements | Energy consumption (/OP) | Computing throughput (OP/s) | Functionality | Single unit speed (OP/s/device) |
|---|---|---|---|---|---|
| MZI mesh[14] | 60 | 30.00fJ | 32T | MVM | 0.53T |
| MRRs[36] | 4 | 0.56pJ | 80G | MVM | 20G |
| PCM[5] | 36 | 2.5pJ | 121.1G | Conv. | 3.36G |
| VCSEL array[37] | 25 | 7.14fJ | 38.4G | MVM | 1.536G |
| OCPU[13] | 4 | 4.84pJ | 128.1G | Conv. | 32G |
| MP (this work) | 1 | 5.94pJ | 73.4G | MVM +Conv. | 73.4G |
| Enhanced MPUs (Fig.10) | 4 | 21.65fJ ($V_\pi = 1V$[31]) | 4.4T (with 110 GBaud[31] and 5 wavelengths) | MVM +Conv. | 1.1T |

In addition, the proposed MP design may introduce a temporal redundancy in certain computational tasks. For instance, when shifting from matrix-vector multiplications to convolution operations, it becomes necessary to temporally reorganize some kernel and image elements to facilitate convolution to maintain the original hardware setup. Such redundancy is resulted from the absence of the reuse of kernels and image data, and therefore leads to increased computational latency for the MP, as detailed in Table 3. Employing strategies like WDM and SDM, as depicted in Fig.10, can alleviate this latency issue. In practical applications, it is crucial to strategically allocate time, space, and wavelength resources within the MP's framework, tailored to the specific demands for computational speed and flexibility pertinent to the scenario at hand.

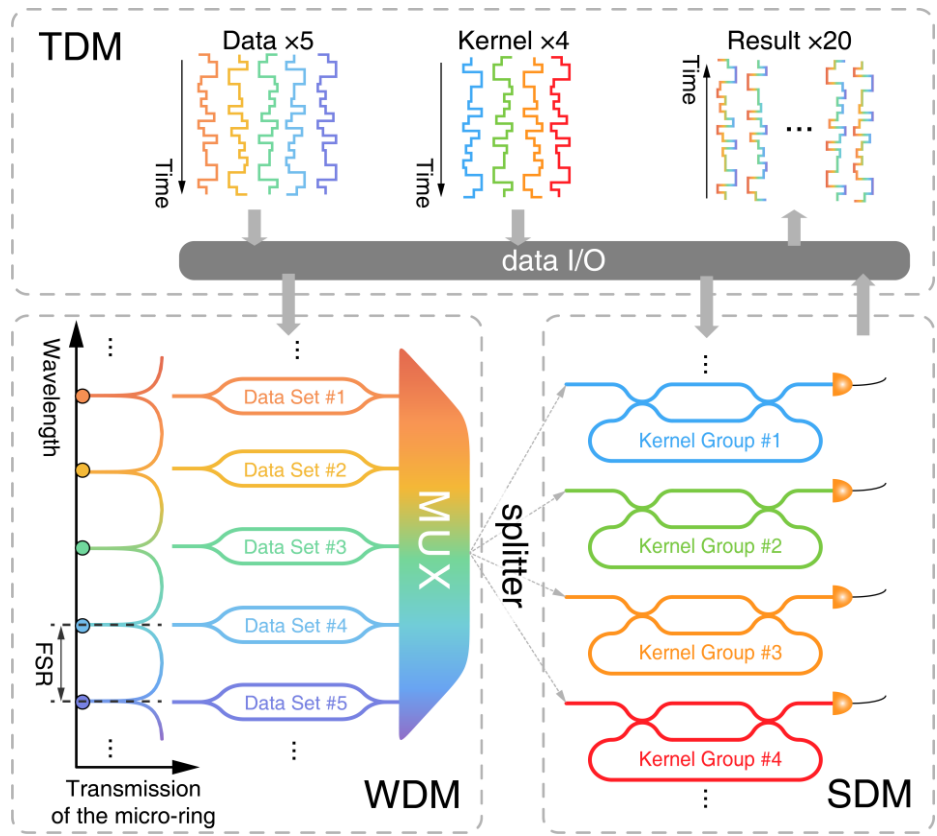

Fig.10 The illustration of enhanced strategy integrating time, space, and wavelength resources.

## Strategic Integration of MP with WDM and SDM

In this study, the MP's computational prowess is primarily derived from the rapid modulation capabilities of electro-optic modulators within a TDM framework. This design allows for a scalable and flexible linear computing architecture on a simple on-chip system. However, as analyzed in the previous section, exclusive reliance on TDM can lead to increased power consumption and latency as computational demands grow. Future research should therefore incorporate various multiplexing dimensions to mitigate these issues in large-scale computations, while maintaining scalability, flexibility and versatility.

Fig.10 showcases an enhanced architecture that integrates multiple wavelengths (WDM strategy) and spatial structures (SDM strategy) with the existing MP. Optical wavelengths are spaced according to the FSR of ring-modulators and theoretically possess uniform modulation and feedback intensities[32]. By encoding different data on these wavelengths and channeling them through the ring-modulators, we can process multiple data sets with a single weight matrix (or convolution kernel), enabling concurrent vector multiplications. This method decreases computational delay or eases the bandwidth demands on input electrical signals, which are in principle benefit from the reuse of data to-be-processed or kernels/weights, thus reduces the frequency of memory access and thus power usage[38]. Furthermore, spatially distributed ring-modulators, each assigned unique weight vectors, enhance computational parallelism, further reducing computational latency or energy consumption.

Nevertheless, the escalation in costs due to multi-wavelength sources and complex electro-optic modulation, coupled with the challenges in fabricating large-scale parallel photonic devices, necessitates a balanced expansion in wavelength and spatial dimensions. Within the limits of the electrical clock frequency, the proposed MP framework, rooted in TDM, can scale up computations on a single device while ensuring a certain computational throughput. In summary, this architecture judiciously utilizes time, wavelength, and spatial resources to merge the adaptability and swift reconfiguration of TDM, with the parallelism of WDM and SDM schemes, and the inherent low-power transmission benefits of optical signals. This synergy offers a novel paradigm for future optical computing frameworks.

# Conclusions

In this paper, we have proposed a novel, flexible, and scalable on-chip MP that utilizes a TFLN ring-modulator. The MP seamlessly performs and transitions between vector multiplication, matrix-vector multiplication, and convolution operations, including both single and multi-kernel convolutions within each computation cycle. This functionality is achieved by adjusting the delay in the ring-modulator's feedback waveguide, coordinating the symbol periods between the external MZM and the ring-modulator, and organizing data symbols temporally. This is accomplished without necessitating alterations to the existing photonic architecture on the chip.

To validate the feasibility of the proposed MP, the convolution operations of the MP,

which include both matrix-vector and vector multiplications, have been experimentally demonstrated by using an on-chip TFLN ring-modulator. An optical convolution layer based on the MP, integrated with an electronic fully connected layer, constitutes a CNN. This CNN has been deployed for a ten-class classification task on the MNIST handwritten digits dataset. In tests involving both single and multi-kernel feature extractions within a single MP convolution, classification accuracies of 88.7% and 88.3% have been achieved, respectively. These results show only slight variations when compared to digital CNNs (3.6% in the single-kernel scenario and 3% in the multi-kernel scenario), illustrating that the MP scaling and function switching maintains computational accuracy without additional hardware, thereby underscoring its remarkable scalability and flexibility.

## Materials and Methods

### Fabrication of the TFLN ring-modulator

The ring-modulator is constructed on a commercial X-cut LNOI wafer. The 1.5-μm-width half-etched LN ridge waveguide is crafted using electron beam lithography and dry etching techniques, resulting in a ridge and slab each 200 nm. The grating couplers, as fiber-chip interfaces, are formed as detailed in [39]. Then, the optical components are encapsulated with an 800-nm $SiO_2$ over-cladding layer with plasma enhanced chemical vapor deposition. Subsequently, totally 1.1-μm-thick Au are deposited as CLTW using two-steps of electron beam evaporations and lift-off. Additionally, 30-μm-depth holes are created in the silicon substrate via dry etching to align the microwave and optical group refractive indices. More details of the fabrication are described in Supplementary Note 4. The fabricated ring-modulator occupies a footprint of 3.4 mm by 0.7 mm, and incurs a total optical loss of approximately 7.15 dB, which includes on-chip insertion losses of about 0.15 dB and fiber-chip coupling losses of roughly 3.5 dB per facet.

### Experiment Setup

As illustrated in Fig.5, in the optical convolution layer, the optical carrier, soured from a DFB laser and at one of the resonant wavelength of the ring in the ring-modulator, is modulated by an electrical data signal $X$ generated from one channel of the arbitrary waveform generator (AWG, Keysight M8199A), through the external MZM. Then the generated optical signal is amplified by an erbium-doped fiber amplifier (EDFA). After passing through the polarization controller (PC), the optical signal is coupled into the on-chip TFLN ring-modulator via the on-chip grating coupler. The ring-modulator, biased at the over-coupled point, is driven by an electrical signal from another channel of the AWG, which is encoded with elements of the coupling vector $W'$ transformed from the weight matrix or the kernels (refer to Section Operating Principle). Two high-bandwidth RF probes (GGB 67A) are employed in this study. One probe delivers the electrical signal to the on-chip ring-modulator, and the other provides a 50Ω termination for impedance matching. Here noted that the symbol rates of the two signals $X$ and

$W'$ should be properly set. The time-domain alignment between signals $X$ and $W'$ is achieved through software-defined channel-wise delay adjustments of the AWG.

After on-chip convolution operations are completed, the optical signal carrying the convolution results is exported from the ring-modulator and then directed to the electrical fully connected layer. Here, a PD (Finisar BPDV2150R) converts the optical signal into electrical one, which is subsequently amplified by RF amplifier, and sampled and by a real-time oscilloscope (Tektronix DSA73304D). The collected data is then transferred to a personal computer for off-line analysis.

## Acknowledgements

This work was supported by National Natural Science Foundation of China (62335019), Innovation Program for Quantum Science and Technology (2021ZD0301401), National Natural Science Foundation of China-Guangdong Joint Fund (U2001601), National Natural Science Foundation of China (62135012), Leading Innovative and Entrepreneur Team Introduction Program of Zhejiang (2021R01001), "Pioneer" and "Leading Goose" R&D Program of Zhejiang (2023C01139).

## Author details

[1]State Key Laboratory of Optoelectronic Materials and Technologies, School of Electronics and Information Technology, Sun Yat-sen University, Guangzhou; 510006, China. [2]Guangdong Provincial Key Laboratory of Optical Information Materials and Technology, South China Academy of Advanced Optoelectronics, South China Normal University, Guangzhou 510006, China. [3] State Key Laboratory of Extreme Photonics and Instrumentation, College of Optical Science and Engineering, International Research Center for Advanced Photonics, Zhejiang University, Hangzhou 310058, China.

## Author contributions

Z.A.D., Z.H.L and J.L. designed, determined and verified the operating principle of MP based on ring-modulator. Z.A.D., C.Y.B., Z.H.L. and Z.H.C. experimentally demonstrate the image convolutions and optic-electronic CNN based on the MP. G.R.F., K.X.C., C.J.G., and L.L. contributed to the fabrication and characterization of the on-chip ring-modulator. J.L., Z.A.D., Z.H.L., C.Y.B., J.Q.L., and S.Y.Y. wrote the manuscript. All authors reviewed the manuscript.

## Conflict of Interest

The authors declare no conflict of interest.

**Supplementary Notes** are available for this paper.

# Supplementary Notes:

# Scalable and Versatile Linear Computation with Minimalistic Photonic Matrix Processor

Zhaoang Deng[†], Zhenhua Li[†], Jie Liu[†,*], Chuyao Bian, Jiaqing Li, Ranfeng Gan, Zihao Chen, Kaixuan Chen, Changjian Guo, Liu Liu* and Siyuan Yu

[†] These authors contributed equally.

*Corresponding author: liujie47@mail.sysu.edu.cn; liuliuopt@zju.edu.cn

# Contents

## Supplementary Note 1: MAC representation with ring-modulator

The transmission rules of the ring-modulator can be derived from the following relationship[1]:

$$\begin{cases} E_{up}(t) = \frac{1}{\sqrt{2}}[E_{in}(t) + jE'_r(t)] \\ E_{down}(t) = \frac{1}{\sqrt{2}}[jE_{in}(t) + E'_r(t)] \\ E'_{up}(t) = E_{up}(t)e^{-\alpha l_m + j\beta_g(l_m + \Delta l(t))} \\ E'_{down}(t) = E_{down}(t)e^{-\alpha l_m + j\beta_g(l_m - \Delta l(t))}, \\ E_{out}(t) = \frac{1}{\sqrt{2}}[E'_{up}(t) + jE'_{down}(t)] \\ E_r(t) = \frac{1}{\sqrt{2}}[jE'_{up}(t) + E'_{down}(t)] \\ E'_r(t + T_s) = E_r(t)e^{-\alpha l_r + j\beta_g l_r} \end{cases} \quad (S1)$$

in which, $\alpha$ and $\beta_g = \frac{2\pi}{\lambda}n_g$ are the loss and the propagation constant of the TFLN waveguides, respectively, $l_m$, $l_r$, and $\Delta l(t)$ is the length of the arms of MZI, the length of the ring, and the differential length of the arms of MZI induced by push-pull electro-optic modulation, respectively, and $T_s$ is the time delay of the ring.

Since the result retained within the micro-ring after each iteration is denoted as $E'_r$, we can derive the update formula for $E'_r$ in each iteration as follows:

$$E'_r(t+T_s) = AB\left[E_{in}(t)\cos\left(\beta_g \Delta l(t)\right) + E'_r(t)\sin\left(\beta_g \Delta l(t)\right)\right], \tag{S2}$$

in which, $A = e^{-\alpha(l_m+l_r)}$ and $B = e^{j\left[\beta_g(l_m+l_r)+\frac{\pi}{2}\right]}$ are the loss and the common phase of the loop of the loop, respectively. Sampling the signal by period of $T_s$, the optical field of Eq.S2 can be denoted as the discrete form:

$$y^{(n)} = E'_{r,n+1} = AB[E_{in,n}c_n + E'_{r,n}s_n], \tag{S3}$$

$$\begin{cases} c_n = \cos(\beta_g \Delta l_n) \\ s_n = \sin(\beta_g \Delta l_n) \end{cases}, \tag{S4}$$

in which the coupling factors $c_n$ and $s_n$ are depended on $\Delta l_n$, or directly the modulation voltage $v_n = \frac{V_\pi}{\pi}\beta_g \Delta l_n$ on the MZI of ring-modulator. The coupling factors satisfies $c_n^2 + s_n^2 = 1$.

We can denote the optical input of the ring-modulator from the external MZM as the data vector:

$$E_{in,n} = \begin{cases} x_n, n = 1, \cdots, N \\ 0, other \end{cases}, \tag{S5}$$

and without loss of generality, assume that:

$$y^{(0)} = E_{r,0} = E'_{r,1} = 0, \tag{S6}$$
$$A = B = 1, \tag{S7}$$

for simplification. In $n^{th}$ ($1 \leq n < N$) iteration,

$$\begin{aligned} y^{(n)} = E'_{r,n+1} &= x_n c_n + y^{(n-1)} s_n \\ &= x_n c_n + [x_{n-1}c_{n-1} + y^{(n-2)}s_{n-1}]s_n \\ &= x_n c_n + x_{n-1}c_{n-1}s_n + [x_{n-2}c_{n-2} + y^{(n-3)}s_{n-2}]s_{n-1}s_n \\ &= \cdots \\ &= x_n c_n + \sum_{i=1}^{n-1} x_i c_i \left(\prod_{k=i+1}^n s_k\right) \end{aligned} \tag{S8}$$

In the last iteration, we need to read out the result from the output port, but not to put back into the loop. So, we need load high voltage on the MZI so that an additional differential phase shift $\pi$ between both arms (i.e., $\beta_g \Delta l_n \to \beta_g \Delta l_n + \pi/2$). Therefore, we can obtain (omitting the loss and common phase from the MZI, only considering the coupling ratio):

$$\begin{aligned} y = y^{(N)} = E_{out,N} &= -x_N s_N + y^{(N-1)} c_N \\ &= -x_N s_N + x_{N-1}c_{N-1}c_N + \sum_{i=1}^{N-2} x_i c_i c_N \left(\prod_{k=i+1}^{N-1} s_k\right). \end{aligned} \tag{S9}$$

Letting the corresponding coupling vector $W' = [\omega'_1, \cdots, \omega'_N]^T$,

$$\omega'_n = \begin{cases} -s_N, n = N \\ c_n, n < N \end{cases}, \tag{S10}$$

i.e., the coupling signal $\omega'_n$ should be temporally aligned to the input signal $x_n$, the results of each step can be rewritten as:

$$y^{(n)} = x_n \omega'_n + \sum_{i=1}^{n-1} x_n \omega'_n \left(\prod_{k=i+1}^N \sqrt{1 - {\omega'_k}^2}\right) = \begin{cases} E_{out,N}, n = N \\ E'_{r,n+1}, n < N \end{cases}. \tag{S11}$$

From the expression, owing to the adjustable coupling factor and the recursive loop of the ring-modulator, in each step there are both multiplication and accumulation conducted, i.e., a MAC operation.

Therefore, to obtain the vector multiplication of data vector X and weight vector W: $y^{(N)} = W^T X = \sum_{i=1}^N x_i \omega_i$, we need to determine proper $W'$ according to $W$ following the relationship:

$$\omega_n = \begin{cases} \omega'_N, n = N \\ \omega'_n \left( \prod_{k=n+1}^{N} \sqrt{1 - {\omega'_k}^2} \right), n < N \end{cases}. \tag{S12}$$

From $\omega'_N$ to $\omega'_1$, the coupling vector $W'$ can be finally determined.

In the experimental demonstration, we only use a PD (with responsivity of μ) to detect the result, obtaining the photocurrent $I = \mu|y|^2$. Then, we get the square root of $I$ and normalize it, resulting in the non-negative $|y|$. It means that if $y$ itself is non-negative (i.e., the vector multiplication of two non-negative vectors), $|y|$ is equal to $y$. But if the multiplied vector contains negative elements, $y$ might be negative and $|y|$ might be wrong result. To the convolution application, the input image data is non-negative, but the kernel may have negative elements, which need additional technique to transform into a non-negative representation, as detailly discussed in Supplementary Note 2.

Back to the assumption $B = 1$, it means that the phase of $E'_r$ should be time-independent and the same as $E_{in}$, so that the amplitudes of $E_{in}$ and $E'_r$ can be completely added up coherently. It holds only when the carrier wavelength lies precisely at the micro-ring's resonant wavelength, demanding the stability of the ring-modulator and the small linewidth of the laser.

## Supplementary Note 2: Representation of negative kernel

If the kernel is expressed as:

$$W = \begin{bmatrix} w_{11} & w_{12} \\ w_{21} & w_{22} \end{bmatrix} = \begin{bmatrix} w^+_{11} & w^+_{12} \\ w^+_{21} & w^+_{22} \end{bmatrix} - \begin{bmatrix} w^-_{11} & w^-_{12} \\ w^-_{21} & w^-_{22} \end{bmatrix} = W^+ - W^-, \tag{S13}$$

where the $W^+ = \begin{bmatrix} w^+_{11} & w^+_{12} \\ w^+_{21} & w^+_{22} \end{bmatrix}$ and $W^- = \begin{bmatrix} w^-_{11} & w^-_{12} \\ w^-_{21} & w^-_{22} \end{bmatrix}$ are the "positive part" and the "negative part" of a real number, respectively, and both with all non-negative entries. To convolute an image X (data matrix, all pixels are non-negative) with the real kernel $W = W^+ - W^-$, ones should sequentially execute the convolutions $X * W^+$ and $X * W^-$ on the MP, then combine two part of convolution results together by differential operation:

$$X * W^+ - X * W^- = X * (W^+ - W^-) = X * W. \tag{S14}$$

## Supplementary Note 3: Read-out and reset of MP

In the MAC operation, when $n < N$, the current MAC result is at the port #3 or #4. To directly read out current result from the output port, one should direct the current result from $E'_r$ to $E_{out}$. Thus, at this read-out $((N + 1)^{th})$ step, there would be the following relationship:

$$E_{out,N+1} = A_m B_m [E_{in,N+1} s_{N+1} + E'_{r,N+1} c_{N+1}] = A_m B_m E'_{r,N+1}, \tag{S15}$$

$$A_m = e^{-\alpha l_m}, B_m = e^{j[\beta_g l_m + \frac{\pi}{2}]}, \tag{S16}$$

This relationship indicates that, if assuming $A_m = B_m = 1$, the coupling factors should be $c_{N+1} = 1$ and $s_{N+1} = 0$ by adjusting $\Delta l_{N+1}$ or $v_{N+1}$.

During the result reading out $E_{out,N+1} = E'_{r,N+1}$, another relationship is expressed as:

$$E_{r,N+1} = E_{in,N+1}c_{N+1} + E'_{r,N+1}s_{N+1} = E_{in,N+1}, \quad (S17)$$

which reveals that, if setting $E_{in,n} = 0, n \geq N+1$, all the energy in the ring would be eliminated, i.e., $E_{r,N+1} = 0$, realizing the reset operation of the MP.

In fact, the last iteration step of vector multiplication introduced in Supplementary Note 1 is a specific example to simultaneously conduct the last MAC operation and read out the final vector multiplication result from port #2. After the result read out, generally additional several iteration steps to reset the device is necessary, to satisfy the assumption of $E_{r,0} = E'_{r,1} = 0$ for the next vector multiplication.

## Supplementary Note 4: Fabrication and characterization of the ring-modulator

The process flow of the proposed ring-modulator[1] was shown in Fig.S1. Initially, on commercially available x-cut LNOI wafer with a 500 um thick silicon substrate layer, a 3 um thick BOX layer, and a 400 nm thick TFLN layer (Fig.S1(a)), the optical components, including two 3 dB direction couplers, two grating couplers, and waveguides were patterned using a MaN-2403 photoresist mask after e-beam lithography (EBPG 5000+) and Mad525 developed. High-energy Ar+ ions bombardment was utilized for half-etched LN circuits (Fig.S1(b)). Subsequently, LN holes structures were fabricated with an AZ5214 mask for etching (Fig.S1(c)). The whole TFLN structures were coated with an 800-nm-thick $SiO_2$ over-cladding layer deposited by plasma enhanced chemical vapor deposition (PECVD) (Fig.S1(d)). Moreover, modulation electrodes were employed by electron beam evaporation in two processes: firstly, a 200 nm thick T-shaped gold electrode was deposited (Fig.S1(e)), followed by a secondary deposition of a 900 nm traveling-wave electrode (Fig.S1(f)). After that, the silicon dioxide cladding and 3 um BOX at specified holes structure locations were fully etched down to the silicon substrate through inductively coupled plasma (ICP) etching techniques. Finally, an isotropic etching of 30 um thick into the silicon substrate completed the chip fabrication (Fig.S1(g)), ensuring the matching of microwave refractive index and optical group refractive index.

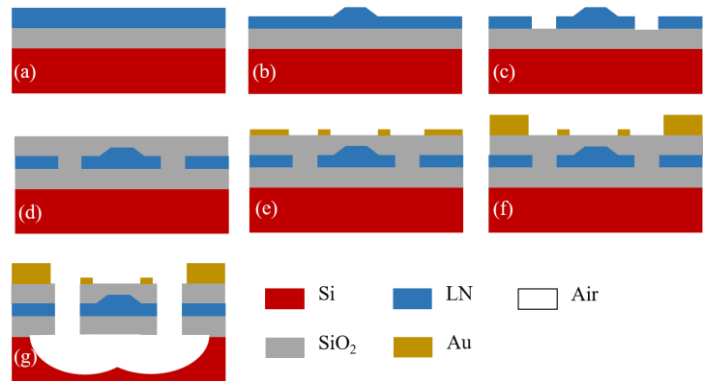

Fig.S1 Process flow of the proposed Micro-ring modulator: (a) Layers on commercial LNOI; (b) LN waveguide fabrication; (c) LN hole etching; (d) $SiO_2$ over-cladding deposition; (e) 200

nm T-shape electrode evaporation; (f) 900 nm travel-wave electrode evaporation; (g) $SiO_2$ holes etching to the silicon substrate and Isotropic silicon etching.

The experimental set-up for characterization of the ring-modulator is shown in Fig.S2(a). For the optical transmission spectrums measurement and the electro-optic transfer curve, a continuous wave (CW) generated by a tunable laser (Keysight 81980A, 1465nm~1575nm) is coupled into the ring-modulator that is biased by a DC voltage source (ITECH IT6302, configured to supply -31V to 30V DC voltage) through a RF probe (GGB 67A, DC~67GHz, G-S-G configuration), and a power meter (Keysight 81634B, C-band) detects the intensity outputted from the ring-modulator.

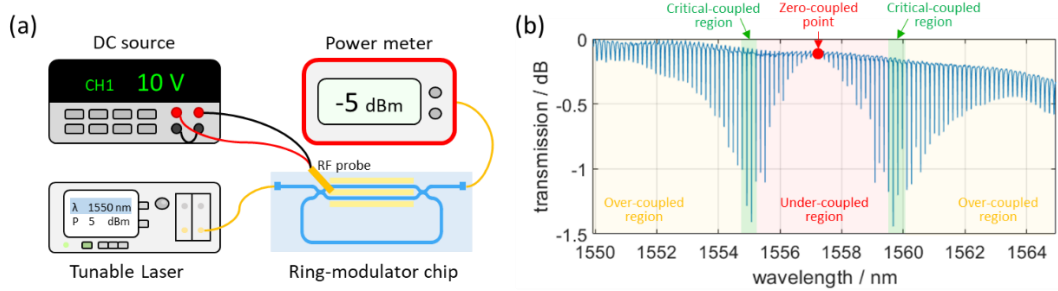

Fig.S2 (a) The experimental set-up for characterization of the ring-modulator. (d) The optical transmission spectrums of the ring-modulator.

For the optical transmission spectrums measurement, the DC source is off and the tunable laser conducts a wavelength sweeping. Thus, the transmission spectrum of each wavelength could be read from the power meter. Fig.S2(b) is the measured spectrum, on which the under/zero-coupled, critical coupled and over-coupled regions are marked. The spectrum is basic on a frequency comb whose dips are the resonant wavelengths of the ring, and the intensity of the comb's dips are modulated by a envelop determined by the coupling ratio of the MZI regarding to different wavelength. The zero-coupled point indicating that the MZI is working in the "bar" state, and no/less energy is coupled into the ring. So, the light directly passes through the device and results in high output intensity. The over-coupled region indicates the MZI is near to the "cross" state, most of the energy is coupled into the ring, but after only one loop, coupled out of the ring since the MZI's "cross" state. So, the output intensity is also high. Between the zero-coupled and over-coupled region, there are the critical coupled region with deep dips, because the coupling ratio and the loss of the ring satisfy a certain relationship, leading to most of the energy undergo destructive interference at the output port[2-4].

Next, for the electro-optic transfer curve measurement, the wavelength should be firstly tuned at one of the resonant wavelengths shown in the transmission spectrum. The DC source conducts a voltage sweeping, and then the transfer curve can be read from the power meter, as shown in Fig.4(c) in the main text. The transfer curve is theoretically analogue to the envelop of the dips in the transmission spectrum, also divided into under/zero-coupled, critical coupled and over-coupled regions, which are also highly relative to the coupling ratio of the MZI. Therefore, $V_\pi$ of the MZI can be determined by reading the half width between two zero-coupled point, which is the voltages needed to switch the MZI between to "bar" and "cross" states (or differential phase shift between two arms of $\pi$). From the electro-optic transfer curve, the bias voltage for MP operation can be only roughly located in the over-coupled region which is relatively flat and wide. To more precisely calibrating the DC bias, one should characterize the coupling ratio of the MZI individually, excluding the influence of the ring. This needs another characterization technique that is introduced in Supplementary Note 6.

## Supplementary Note 5: Analysis on instability of MP

In the iterative computations of the MP, even small experimental errors can accumulate to an extent that significantly impacts the experimental results. Among these errors, phase instability of the external MZM and the micro-ring is the most pronounced. The direct impact is that when the phase of the input optical signal $E_{in}$ and the ring $E'_r$ are no longer fixed, the linear superposition of optical fields $E_{in}$ and $E'_r$ may not lead to the direct sum $E_{in} + E'_r$. With these kinds of phase errors, Eq.2 of the main text can be rewritten as:

$$y = \sum_{i=1}^{N} x_i\, e^{j\phi_i} \omega_i e^{j\theta_i}, \tag{S18}$$

Where $\phi_i$ and $\theta_i$ are the phases attached on the signal $x_i$ and the MAC result of each steps. Hence, it is imperative to minimize the phase variations of the external modulator and the micro-ring to guarantee their phase stability, so that $\phi_i = \phi$ and $\theta_i = \theta$ which could be after PD detection.

To achieve pure amplitude modulation to encode voltage signals into optical signals, it is imperative to employ the chirp-free push-pull modulation to minimize the phase variations, i.e., $\phi_i = \phi$. We have selected the Fujitsu FTM7961EX and Thorlab LN05S as our external modulators, both of which employ push-pull dual-arm modulation, suitable for pure intensity modulation with phase either 0 or π, standing for positive or negative number, respectively.

Refer to Supplementary Note 1, if the carrier wavelength precisely locates at the resonant wavelength of the ring, the condition $\theta_i = \theta$ is satisfied. However, the resonant wavelengths or the transmission spectrum of the ring-modulator are highly sensitive to the external environment. To assess the impact of the external environment on the micro-ring, Fig.S3(a-c) presents optical transmission spectra from three separate experiments initiated at different times (the intervals among the three experiments are less than one hour), where still exhibit several aspects of discrepancies. Nevertheless, some experimental strategies could be implemented to mitigate the instability.

- Firstly, variations are observed in the intensity of light in the flat regions away from the dips, possibly due to the mechanical and polarization fluctuations in fiber-chip coupling losses caused by fiber jitter. These variations could be mitigated by advanced fiber-chip coupling[5] and on-chip packaging [6-12] technologies.
- Secondly, depicted in Fig.S3(d), there are slight variations in the positions of the dips, attributed to micro-ring dips random drift caused by environmental changing like temperature perturbations. Although the drift wavelength does not exceed 0.01 nm, it may still introduce errors during MAC operations. These variations could be mitigated by applying temperature control[13-16].

- Lastly, also as depicted in Fig.S3(d), the intensity/depth of the corresponding dips are different in the three experiments. This variation possibly stemming from bias voltage drifting on the MZI of the ring-modulator, whose characterization and calibration will be further discussed in Supplementary Note 6.

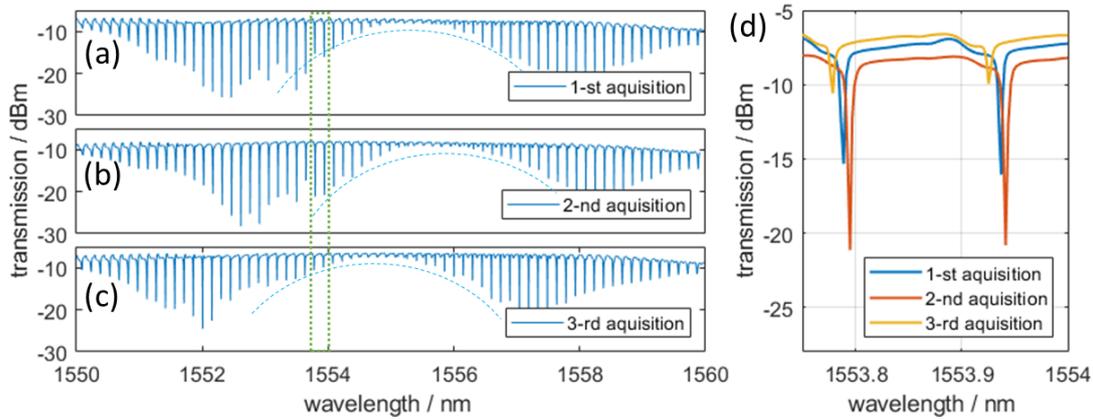

Fig.S3 (a-c) Optical transmission spectrum of the ring-modulator chip in three acquisitions. (d) The zoom-in spectrum of (a-c) near 1554 nm (dashed region in (a-c)).

Although the instability of MP could be addressed to some extent by the measures, the residual signal distortions and low system signal-to-noise ratio (SNR) still contribute the performance discrepancy. Although CNNs inherently exhibit a certain level of error tolerance to limited bit precision[17], the further enhancements on device stability and system SNR by utilizing strategies on reducing device losses[18] and coupling loss[5], and equipped with low-noise trans-impedance amplifiers (TIA)[19] and so on, could significantly bolster performance.

## Supplementary Note 6: Characterization and real-time calibration of the MZI of MP

Since the chip is greatly influenced by external environmental factors, the DC bias point of the MZI of the ring-modulator may continuously drift (trend to increase) during the experiment, as exhibited in Fig.S4, becoming a key aspect degrading the stability of the experiments. This phenomenon has been found and solved in previous researches[20-23]. It can also be seen from Fig.S4 that the DC bias point drifts more rapidly within the first 10-15 minutes, and then gradually slows down, eventually stabilizing around 8.5V. Therefore, in practice, it is necessary to alternately characterize and calibrate the MZI of the ring-modulator multiple times until their parameters gradually stabilize.

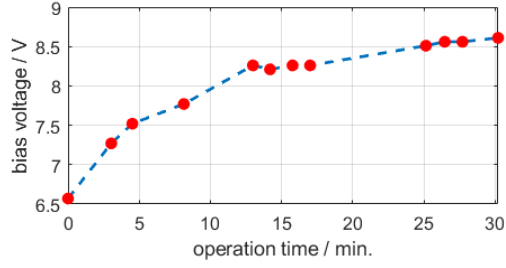

Fig.S4 The variation curve of the drifting DC operating point of the MZI of the ring-modulator.

For the ring-modulator, although the $V_\pi$ can be found via the bias sweep method as introduced in Supplementary Note 4, the basic positioning of actual bias point is quite different from that for traditional MZM. Since the MZI of the ring-modulator is tightly connected to the micro-ring, we use a pulse (delta function) generated by the external MZM to evaluate the operating point of the micro-ring. The pulse width is shorter than the delay of the ring, which means that the intensity of the output pulse series mirrors the impulse response of the ring-modulator. Specifically, the intensity of the first pulse in the series directly reflects the instantaneous transmission from port #1 to #2, and thus the coupling factor of the MZI of the ring-modulator.

Based on the technique to obtain instantaneous transmission of the MP, ones can find the electro-optic transfer curve that is consist of a series of the instantaneous transmission by scanning the modulating voltage on the MZI, as shown in Fig.S5. Specifically, the lowest point of the curve is the over-coupled state of the ring-modulator, where we read out the MAC results and reset the MP, according to Supplementary Note 3. Therefore, we should always set the DC bias voltage of the MZI of the ring-modulator at this over -coupled point, which means that when we set the electrical AC input signal to zero, we can read out the MAC result of the MP and reset the MP. According to the characterization of Fig.S5, the DC bias voltage is 4.12V.

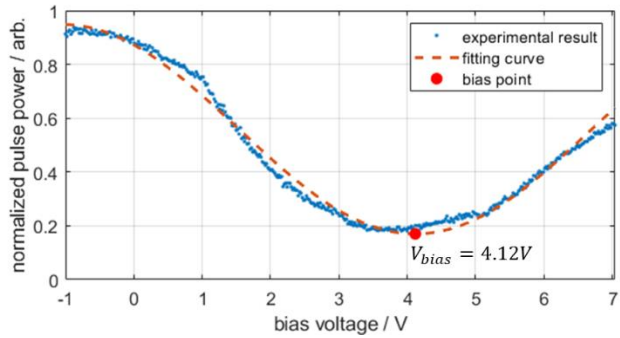

Fig.S5 Electro-optic transfer curve of the MZI in the ring-modulator.

## Supplementary Note 7: Convolution error and bit precision

To evaluate the performance of the image edge extraction of the proposed LPU system, the RMSE and bit precision between the convolution result with the digital computer and the proposed MP system is calculated.

RMSE is a widely used metric for quantifying the difference between values predicted by a model (the value calculated by the computer) and the actual values observed from the system. As one of the most common objective criterion used to evaluate the performance of image processing[24], RMSE can be calculated as follows:

$$\text{RMSE} = \sqrt{\frac{1}{KMN}\sum_{k=1}^{K}\sum_{i=1}^{M}\sum_{j=1}^{N}e_{ij}^{(k)\,2}} \tag{S19}$$

where $e_{ij}^{(k)} = a_{ij}^{(k)} - b_{ij}^{(k)}$ is the pixel error, $a_{ij}^{(k)}$, $b_{ij}^{(k)}$ are the pixel values of the $i^{th}$ row and the $j^{th}$ column from the feature images (the convolution results) of the $k^{th}$ image with the experimental MP system and the digital computing, respectively, $M$ and $N$ are the number of rows and columns of the feature image respectively, and $K$ are the number of image.

The bit precision $N_b$ is calculated as follows:

$$N_b = \log_2\left(\frac{\mu_{max} - \mu_{min}}{\sigma}\right) \tag{S20}$$

Where $\mu_{max}$ and $\mu_{min}$ are the maximum and minimum values of the output, respectively. $\sigma = \sqrt{\frac{1}{KMN}\sum_{k=1}^{K}\sum_{i=1}^{M}\sum_{j=1}^{N}\left(e_{ij}^{(k)} - \bar{e}\right)^2}$ is the standard deviation of $e_{ij}^{(k)}$, the errors between the experimental output and expected output, and $\bar{e}$ is the mean of $e_{ij}^{(k)}$. In this case, the feature images of both experimental and digital systems are normalized, therefore it is obvious that $\mu_{max} = 1$ and $\mu_{min} = -1$. The bit precision $N_b$ of the experimental result mainly depends on the stability of the micro-ring and the MZI of the ring-modulator chip, which has no concern with the specific modulating value of the convolution kernel.

Therefore, for single-kernel convolution, we choose 20 feature images (with size of 27×27 pixels) extracted by 4 kernels shown in Fig.7 of the main text to calculate the RMSE and bit precision. The resulted feature images totally contain 20×27×27=14580 pixels, whose normalized experimental are compared with the ideal results in Fig.S6, and utilized to figure out the RMSE of 0.0699 by Eq.S19. The inset figure in Fig.S6 shows the distribution of errors $e_{ij}$, revealing the standard deviation $\sigma = 0.0698$. Thus, according to Eq.S20, the bit precision $N_b = 4.84\ bit$. The RMSE of dual-kernel convolution of 0.1494 and bit precision of 3.74 bit are derived from similar computations.

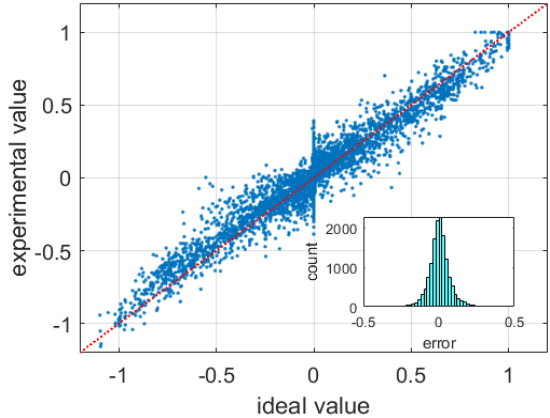

Fig.S6 Comparison between experimental results and ideal results of 14580 pixels of the feature images. The inset shows the distribution of the error $e_{ij}^{(k)}$.

## Supplementary Note 8: Definition and Calculation of computing metrics

(1) Computation throughput, single unit speed and image processing speed

In this paper, the computation throughput is defined as the number of computing operations completed per second by the computing system. In our MP, each symbol period implements a multiply operation and an accumulate operation (Note that 1 MAC = 1 multiplication + 1 accumulation = 2 OP). Therefore, the computing throughput in the MP is $36.7\text{GBaud} \times 2 = 73.4\text{ GOP/s}$ for single-kernel convolution and $36.7\text{GBaud} \times 2 = 73.4\text{GOP/s}$ for dual-kernel convolution. In the enhanced MP, 5-wavelength WDM and 4-ring-modulator SDM extend the computing throughput of MP by 20 times. With a baud rate of 110GBaud[25], the computing throughput is $110\text{GBaud} \times 20 \times 2 = 4.4\text{TOP/s}$.

The single unit speed is simply defined as the computing throughput per computing unit. So, the single unit speed is $73.4\text{GOP/s/unit}$ and $\frac{4.4\text{TOP/s}}{4} = 1.1\text{TOP/s /unit}$ for the demonstrated MP and the enhanced MP, respectively.

Based on the computation throughput, for practical image processing applications, the image processing speed is also an important metric (refer to Table 1 and 2 in the main text). The MP requires $(28 - 1) \times (28 - 1) = 729$ MAC operations for a $28 \times 28$ handwritten digit image convoluting with a $2 \times 2$ kernel. Each MAC operation requires a minimum of $4 + 1 = 5$ symbol periods (with at least one additional period for reset, refer to Supplementary Note 3). Given that the convolution kernel contains negative numbers, it is necessary to subtract the results of two positive convolution kernel operations to obtain the final convolution result (refer to Supplementary Note 2). Consequently, the MP can process handwritten digit images at a rate of $\frac{18.35\text{Gbaud/s}}{5 \times 27 \times 27 \times 2} = 2.52M\ image/s$ for single-kernel convolution and double rate of $5.04M\ image/s$ for dual-kernel convolution. For comparison, the convolution on a digital computer carried out by a MATLAB program spends about 7 microseconds, or a processing rate of $0.14M\ image/s$, read by the software profiler.

(2) Number of programmable elements

The number of programmable elements in an individual computing unit often indicates the ability to handle arbitrary or universal operations. On the other hand, it also reveals the fabrication and controlling complexity. In the cases of similar computing scale, more programmable elements mean that more costs the architecture needs to pay to expand the computing scale, including the manufacture of larger arrays, and the complexity of configuration and calibration. In our demonstrated MP, only one ring-modulator is employed but great scalability due to the TDM strategy. And in the enhanced MP framework, an array with 4 ring-modulators is expand the computational capacity by 20 times, with the assistance of WDM strategy.

(3) Energy consumption

In our work, the energy consumption is the energy consumed every operation (including multiplications and accumulations), which consists of two parts: the optical energy mainly from the laser $E_{laser}$, and the electrical energy for modulations $P_{mod.}$, and can be written as:

$$E_{total} = E_{laser} + E_{mod.}. \tag{S21}$$

In our experiments, we use the $P_{laser} = 10mW$ laser source and the baud rate is $B = 36.7 GBaud$, thus the optical energy is:

$$E_{optical} = \frac{P_{laser}}{B} = 0.27 pJ/MAC = 0.135 pJ/OP. \tag{S22}$$

The electrical modulation energy derives from the driving energy for the external MZM $E_{ext}$ and the ring-modulator $E_{RM}$. In the dual-kernel convolution, the driving peak-to-peak voltage are $V_{ext} = 3.5V$ and $V_{RM} = 4.25V$ and the signal baud rate are $B_{ext} = 18.35 GBaud$ and $B_{RM} = 36.7 GBaud$, for the external MZM and the ring-modulator, respectively. Therefore, the modulation energy is[26]:

$$E_{mod} = E_{ext} + E_{RM} = \frac{V_{ext}^2}{2RB_{ext}} + \frac{V_{RM}^2}{2RB_{RM}} = 11.60 pJ/MAC = 5.8 pJ/OP. \tag{S23}$$

Thus, the total energy is:

$$E_{total} = E_{laser} + E_{mod.} = 5.935 pJ/OP. \tag{S24}$$

In the enhanced framework, WDM with 5 wavelengths and SDM with 4 ring-modulators are utilized, which need 5 light source and 9 drivers, and yields 20 vector multiplication result in parallel. Assuming that:

(i) $V_{mod.} = 1V$ and $B = 110\ GBaud$ which could be obtained from the-state-of-art work shows the TFLN coherent modulators[25],

(ii) $P_{laser} = 1mW$ under higher-efficiency PDs and TIAs, and lower-loss optical devices and package (even much lower laser power is needed for optical computing[27]),

we can calculate the energy consumption:

$$E_{optical} = 5 \times \frac{P_{laser}}{B}/20 = 2.3 fJ/MAC = 1.15 fJ/OP, \tag{S25}$$

$$E_{mod} = 9 \times \frac{V_{mod.}^2}{2RB}/20 = 40.9 fJ/MAC = 20.5 fJ/OP, \tag{S26}$$

$$E_{enhanced} = P_{optical} + P_{mod.} = 21.65 fJ/OP. \tag{S27}$$

## Supplementary References